\documentclass[11pt]{article} % 11pt for readability
\usepackage[utf8]{inputenc}
\usepackage[T1]{fontenc}
\usepackage{amsmath}
\usepackage[margin=1in]{geometry} % margin
\usepackage{hyperref}
\usepackage{enumitem}
\usepackage{ragged2e}     % \justifying
\usepackage{setspace}     % line spacing

\title{SETI Post-Detection Futures: Directions for Technosignature Research and Readiness}
\author{}
\date{}

\begin{document}
\maketitle
\justifying

\begin{center}
\small
Kate Genevieve* (Astro Ecologies Inst.), 
Andjelka B. Kovacevic (Univ. of Belgrade), 
John Elliott, 
Martin Dominik, 
Emily Finer (Univ. of St Andrews), 
Kathryn Denning (York University), 
Chelsea Haramia (Univ. of Bonn), 
George Profitiliotis (Blue Marble Space), 
Carol A. Oliver (UNSW Sydney), 
Anamaria Berea (George Mason Univ.), 
Arik Kershenbaum (Univ. of Cambridge), 
Daliah Bibas (Vrije Universiteit Brussels), 
Hannah Little (Univ. of Liverpool), 
William H. Edmondson (Univ. of Birmingham), 
Pauli Laine (NoRCEL)\\[1ex]
Corresponding author: \texttt{kategenevieve@astroecologies.org}\\
Written on behalf of the SETI Post-Detection Hub, University of St Andrews. All authors are Hub members. Full list of 46 members/endorsers: \url{seti.wp.st-andrews.ac.uk/researchers}.
\end{center}

\section{Introduction}

This white paper highlights the work that is needed to anticipate the challenges and societal impacts of a possible technosignature detection. We recommend practical steps to strengthen NASA's astrobiology agenda, guided by the existing interdisciplinary framework of the SETI PostDetection Hub (est. 2022) at the University of St Andrews (Elliot et al./ 2023), which emphasizes comprehensive preparedness across science, society, governance, and communication. NASA can significantly enhance readiness by supporting deep interdisciplinary integration, funding SETI post-detection research infrastructure, and cultivating international collaboration. We outline six key dimensions of readiness-directed evidence-based research: cross-divisional methodologies, humanities and social sciences integration, communication, strategic foresight, and development of resilient global infrastructures.

\section{The Imperative for Strategic Preparedness}

NASA's Center for Life Detection (2018), the Biosignatures Standards of Evidence Workshop (2021), and the Astrobiology Federation (2024), demonstrate a growing commitment to the value of cross-divisional work in astrobiology. Strategic documents such as the Decadal Survey (NASA, 2024) emphasize ``the value of\ldots international partnerships,'' in alignment with the United Nations' goals for global collaboration. Nevertheless, a technosignature discovery could emerge in any realm of astronomy research, whether explicitly designed for biosignatures, technosignatures, exoplanets or other astrophysics investigations conducted by NASA. Such a discovery would instigate unusual challenges, requiring exceptional interdisciplinary integration and advance preparation.

A technosignature detection will trigger a complex global process shaped by uncertainty, misinformation, and multiple ideological stakeholders (Denning et al.\ 2019). Past cases, e.g.\ Jocelyn Bell Burnell's 1967 discovery of pulsars (Penny 2013), highlighted the importance of rigorous, transparent protocols for managing scientific findings. Early guidelines, including the IAA's 1989 ``Declaration of Principles Concerning Activities Following the Detection of Extraterrestrial Intelligence''  set foundational standards of scientific rigor and transparency, yet predate the internet and contemporary technologies. Updated protocols from the 2010s could also not account for the complexity of rapid global media dissemination, and fresh revision efforts are currently underway (Oliver et al.\ 2023; Tennen et al.\ 2024).

\section{Priorities}

Recognizing key gaps, the SETI Post-Detection Hub is developing a comprehensive, long-term institutional response, integrating science, policy, and communication strategies to prepare for future discoveries and devise anticipatory governance and post-detection scenarios. We recommend strategic investment in six key dimensions: 
\begin{enumerate}[label=(\arabic*)]
    \item advanced detection technologies and methodologies,
    \item innovative ``Other Minds'' paradigms beyond anthropocentric assumptions,
    \item humanities and social sciences for interpreting public response, ethics, and governance,
    \item strategic foresight, scenario planning, and anticipatory governance,
    \item public communication strategies and research on communication, and
    \item global coordination on infrastructures for resilience and sustained preparedness.
\end{enumerate}

\subsection{Emerging Scientific Techniques in Technosignature Searches}

Ongoing advances in computational methods will continue to reshape technosignature searches over the next decade, necessitating cross-sector collaboration to integrate multi-modal detection across instruments and AI-driven analysis (Sheikh et al. 2021). As capabilities improve, international efforts (e.g.\ SKA, ALMA and FAST), alongside NASA's JWST, are transforming signal analysis, enabling advancements in subtle anomaly detection (Schwieterman 2024), and leveraging LLMs for deep archival data exploration (Mason et al. 2024). Future transformations can be shaped by the Vera C. Rubin Observatory and NASA's planned observatories (e.g.\ Nancy Grace Roman Space Telescope, HWO). With its capacity for longterm investment, NASA is well-positioned to drive innovation in next-generation sensing technologies.

Our major proposition is that NASA ambitiously boost the methodological potential for technosignatures with a multi-node, networked sensing strategy capable of identifying more complex structures and systemic anomalies in cosmic data. Just as LIGO revolutionized gravitational wave astronomy, an ambitious multi-node SETI sensing network could transform our approach to detecting and interpreting complex extraterrestrial phenomena by correlating multi-wavelength data in real time and identifying cross-signal patterns. We encourage NASA to intervene at the cutting edge, supporting lunar farside low-frequency radio telescopes, Mars exogeoconservation in the search for a second genesis of life in our solar system (Fletcher 2024), and investigating theoretical SETI (e.g.\ Hippke's work on quantum parallelism for weak signal detection (Hippke 2021)) through proactively collaborating with technology industries. Once-speculative ideas are becoming feasible, as seen in climate modelling and space weather prediction (Sripat 2024). Towards this, we recommend that NASA Astrobiology strengthen technical exchanges, industry partnerships, and multi-use instrumentation strategies to maximize scientific benefit, and enable technosignature research to inform and shape the fields of astrobiology and technology more broadly. NASA can powerfully contribute with:

\begin{itemize}
    \item Cross-divisional collaboration on technosignatures and interdisciplinary team infrastructures, and funding where appropriate, to unlock next-generation detection capabilities.
\end{itemize}

\subsection{Expanding Detection Paradigms through ``Other Minds'' Research}

To move beyond Earth-centric assumptions in technosignature detection, future research can engage computational methods from non-human cognition studies. Considering how to investigate extraterrestrial cognitive, sensory, or cybernetic architectures benefits from approaches drawing from biology and the material sciences, neuroscience and philosophy of mind, and others (Elliott 2015). Techniques from bioacoustics, machine learning and quantum computing offer significant cross-disciplinary insights. For example, bioacoustic techniques used to analyze complex animal communications, such as whale song (Sharma 2024, Arnon 2025), may reveal novel ways to recognize patterns, syntax, and symbolic structures in potential extraterrestrial signals. Whilst research on avian magnetoreception, potentially linked to quantum entanglement in cryptochrome proteins, suggest alternative models for navigation and communication. Collaborations between neuroscientists and philosophers continue to refine conceptual frameworks, exploring how consciousness and signal detection might function in alien contexts. Computational modeling, drawing on thermodynamics, entropy minimization and the free energy principle, may offer further ways to explore life's structuring principles across cosmic distances. NASA can play a key role in advancing this interdisciplinary approach by:

\begin{itemize}
    \item Funding interdisciplinary research on non-human communication and cognition by hosting meetings and funded collaborations that drive cross-sector innovation, developing novel modeling and detection paradigms, and establishing long-term technical infrastructures integrating emerging technologies and bio-inspired methodologies.
\end{itemize}

\subsection{Integrating Humanities and Social Sciences for Human Response}

Collaborative and potential funding opportunities with international researchers include the social sciences and humanities, which have been consistently acknowledged since the Brookings Report (Michael 1961, Dews 2014) and since the earliest days of the search for life at NASA (Dick 2013, 2020; Charbonneau 2024) and continues to be on the NASA Astrobiology agenda (e.g., Denning et al.\ 2024). This research and policy work needs to happen before a confirmed discovery is made. Many relevant research methods can contribute. For example:

\textbf{Modeling responses:} Emerging methods from computational social science, including surveys, large-scale sentiment analysis and social media sampling offer predictive models for modeling public and institutional reactions to detection. More speculatively, to consider non-human and nonbiological intelligence, it is possible to extrapolate multi-dimensional impact models for the discovery of extraterrestrial intelligence (Vidal 2015).

\textbf{Science fiction} (e.g.\ film, literature) forms a storytelling reservoir of human responses and scenarios, shaping public perception, and can be synthesized to inform decision-making and ethical engagement (Baxter 2011, Wright \& OmanReagan 2018). Science fiction models diverse first-contact perspectives, societal responses, and enables a valuable research strategy to approach epistemological challenges in interpreting the unknown (Puranam 2024). The SETI Post-Detection Hub's Imagination and Story-telling Working Group specifically responds to this and analyses global science fiction for SETI insights.

\textbf{Ethics and Governance:} Formal foundational theoretical work that defines key concepts like life, intelligence, signals, and communication, alongside refining understandings of the working of analogy, metaphor, and scenarios work, is crucial to contextualize robust post-detection frameworks (Haramia, 2024). Practically, law and data privacy, AI-human collaboration in signal verification, and responsible communication protocols must underpin transparency, trust, and ethical integrity in post-detection (Tennen et al.\ 2024). Further recommendations from the community include:

\begin{itemize}
    \item Funding research on the psychological, social, and global dynamics of post-detection scenarios, including computational social science and digital humanities.
    \item Working with science fiction analysis as an anticipatory tool to inform public engagement strategies, mapping science fiction's influence on public perceptions.
    \item Galvanising NASA's leadership role in guiding international space agreements towards key issues, such as protecting lunar far-side's radio quiet zone for astronomy (Michaud 2024).
\end{itemize}

\subsection{Futures Methods and Scenarios}

The SETI Post-Detection Toolkit 2050 (Profitiliotis et al.\ 2025) is a resource for the SETI and technosignature community integrating contextual future scenarios and alternative detection situations to be used for anticipating post-detection outcomes across various interest groups. This ties to SETI's strategic foresight legacy, and also to independent futures work like Cultures Of The Imagination (COTI) (Funaro 1994) that engaged NASA in speculative exploration. The Toolkit provides crucial definitional groundwork for developing comparative post-detection scenarios by systematically integrating broad expertise to anticipate societal responses, and also---importantly---rigorously modeling and stress-testing strategic plans under novel risks and opportunities. This approach ensures that post-detection strategies continuously evolve alongside scientific and technological advancements. NASA is encouraged to engage with the toolkit for crafting robust plans for navigating post-detection, to aid iteratively testing and refining SETI frameworks under conditions of deep uncertainty. NASA's support would be highly valuable in:

\begin{itemize}
    \item Developing and validating anticipatory institutional practices through scenario planning, stress-testing, and participatory simulation, with stakeholder engagement in the co-design of governance and decision-making strategies, with an emphasis on participation across scientists, agency leaders in astrobiology institutions, and society stakeholders and public.
    \item Funding streams for strategic foresight studies on post-detection and anticipatory governance, with community buy-in to translate the work to the space community, establishing dedicated platforms for the publication and dissemination of such research, addressing the current absence of appropriate venues.
\end{itemize}

\subsection{Communication and Imagination}

In rapidly evolving media landscapes, research is essential on public communication and how storytelling bridges gaps in science communication (Oliver et al. 2023). Art-science research also offers unique ways to investigate how communication is interpreted, providing practical insights into social dynamics, meaning-making, and public response. NASA's own legacy, through the Voyager mission and the Golden Record, demonstrated how crafting messages for unknown recipients can reveal fundamental differences in interpretation. Today, universal message design projects extend this work, and performance-based research, such as Daniela de Paulis' \textit{A Sign in Space} (2019--ongoing), serves as a live testbed, simulating detection scenarios to examine real-time public engagement and inform post-detection protocols. NASA could further support by:

\begin{itemize}
    \item Funding interdisciplinary and participatory research, including citizen science and art-science collaborations, to enhance public communication strategies and anticipatory capacity.
    \item Expanding NASA's Artist-in-Residence (AIR) program to involve artists, philosophers, technologists, and science-fiction creators actively in issues of post-detection.
\end{itemize}

\subsection{Building Resilient Global Infrastructure}

Major astrobiology discoveries require real-time coordination, governance, and crisis response. Without a Post-Detection SETI Hub, NASA risks a gap in the system, akin to a Moon landing without astronaut retrieval or public engagement. Just as NASA developed aerospace emergency protocols for Apollo and post-landing (PL) procedures such as quarantine, scientific analysis, and public communication, SETI post-detection must establish adaptive frameworks ready for ``live'' situations that integrate scientific, diplomatic, and societal expertise to manage verification, communication risks, and global response. Insights from Apollo missions and aerospace crisis units, which demonstrated the value of structured emergency protocols, high-stakes coordination, and adaptive decision-making, can actively inform post-detection planning. NASA can invest in training teams and sustaining preparedness across the ecosystem, particularly the next generation, recognising that a potential technosignatures discovery will subject scientists to immediate public scrutiny and a demand for answers. NASA's role may include:

\begin{itemize}
    \item Cultivating a global post-detection response network, aligning with crisis management models and NASA's communication protocols to sustain operational readiness across time zones.
    \item Defining expert standards for technological resilience, digital governance, secure knowledge sharing and cryptographic standards for data protection in the event of discovery.
    \item Investing in capacity-building around preparedness and communication, with targeted training for early-career researchers (ECRs).
\end{itemize}

\section{Conclusions: Strategic Objectives and Recommendations}

NASA Astrobiology can establish adaptive operational frameworks for real-time global coordination, ensuring robust communication, governance, and technological resilience in the event of a detection. We recommend that NASA embeds interdisciplinary expertise into ongoing research, policy development, and public engagement, and engages with international SETI PostDetection efforts. Specifically, NASA can take action by championing interdisciplinary excellence in technosignatures, integrating science, governance, social sciences and humanities to shape evidence-based first-response strategies within its astrobiology agenda. It can establish long-term infrastructure through sustained funding, dedicated platforms for publishing and archiving, and workshops that embed technosignatures and post-detection research into education and public communication, framing these fields as central to the pragmatic exploration of life beyond Earth. Additionally, NASA can strengthen international collaboration by developing global partnerships and shared governance structures, readiness frameworks, and coordinated public engagement strategies to anticipate future detection events. By investing in long-term infrastructure, interdisciplinary research, and global collaborations, NASA Astrobiology can significantly advance preparedness, positioning itself as a leader in post-detection readiness and strategic response.

\section*{Key References}

\begin{enumerate}[label={[\arabic*]}]
    \item Elliott et al. (2023) The SETI Post-Detection Hub: Preparing for Discovery. IAC-23,A4,2,2,x79324
    \item Denning et al. (2019) SETI and Post-Detection: Towards a New Research Roadmap. In IAC-19-A4.2.4.x52210 IAC-19,A4,2,4,x52210
    \item IAA SETI Permanent Committee. Declaration of Principles Concerning Activities Following the Detection of Extraterrestrial Intelligence, \url{https://iaaseti.org/en/protocols/} 1989.
    \item K. Denning. Social sciences and astrobiology: Some orientation and future directions, 2024. Workshop briefing for NASA Astrobiology Virtual Workshop Feb/Mar 2024.
    \item Tennen et al. (2024), The Future of the SETI Post-Detection Protocols. IAC-24-A4.2.84417
    \item Profitiliotis et al. (2025), \textit{SETI Post-Detection Toolkit 2050}, is the second-stage release of the SETI Post-Detection work, following Genevieve et al. (2024), “Plurality in Post-Detection Scenarios,” 53rd IAC, Milan.
    \item Oliver et al. (2023). Revising the SETI Post-Detection Protocols for the 2020s and Beyond. IAC-23,A4.2.3.79618
\end{enumerate}

\end{document}